# Agent-based Vs Agent-less Sandbox for Dynamic Behavioral Analysis

Muhammad Ali
*School of Computing, Electronics and Mathematics,*
*University of Plymouth*
Plymouth, UK
muhammad.ali@plymouth.ac.uk

Stavros Shiaeles
*School of Computing, Electronics and Mathematics,*
*University of Plymouth*
Plymouth, UK
stavros.shiaeles@plymouth.ac.uk

Bogdan V Ghita
*School of Computing, Electronics and Mathematics,*
*University of Plymouth*
Plymouth, UK
bogdan.ghita@plymouth.ac.uk

Maria Papadaki
*School of Computing, Electronics and Mathematics,*
*University of Plymouth*
Plymouth, UK
maria.papadaki@plymouth.ac.uk

*Abstract*—Malicious software is detected and classified by either static analysis or dynamic analysis. In static analysis, malware samples are reverse engineered and analyzed so that signatures of malware can be constructed. These techniques can be easily thwarted through polymorphic, metamorphic malware, obfuscation and packing techniques, whereas in dynamic analysis malware samples are executed in a controlled environment using the sandboxing technique, in order to model the behavior of malware. In this paper, we have analyzed Petya, Spyeye, VolatileCedar, PAFISH etc. through Agent-based and Agentless dynamic sandbox systems in order to investigate and benchmark their efficiency in advanced malware detection.

*Keywords-Malware detection; Static analysis, Dynamic analysis, Cuckoo, VMRay*

## I. INTRODUCTION

Malicious software also referred as a "Malware" in the cyber security domain [1] saw a significant increase in number of variants. The Internet security report of Symantec concluded that 350 million malware variants were developed in 2016 [2], while Panda Security indicated that, on average, 160,000 new malware programs appeared every day in 2013 [3]. The significant growth in malware variants is due to the ubiquitous nature of computer systems and networks as well as the potential financial benefits of gaining access to computer systems and/or their data.

Although several methods based on shallow architecture have been used in the past by AV companies to detect malware variants, which are either signature based [4], or heuristic-based [5], authors of malware often chose to implement robust, stealthy and sophisticated methods, such as obfuscation, polymorphic and metamorphic mechanism, in order to impede commercial antivirus companies. To cope with the thousands of new malware samples that are discovered every day, security companies and analysts use different techniques. Malware samples are analyzed usually through static and dynamic analysis. In the static analysis, the executable binary of malware sample is analyzed without executing it. This technique is widely used by AV companies to detect malware as this technique uses the concept of pattern recognition and detect malware signature by using a common sequence of bytes in the binary code of a malware. This technique is fast and does not need any controlled environment to run samples, but it can be easily thwarted through obfuscation, packing, and code rearranging techniques and it fails to cover the zero-day malware attack [6]. In dynamic analysis malware, samples are executed and monitored in the controlled environment in order to understand the runtime behavior. This approach is computing intensive as it requires running the malware samples in an isolated sandbox environment in order to obtain artifacts and features, but it tends to have higher accuracy in characterizing malware samples. This technique is agnostic to the underlying code and can easily bypass code obfuscation and polymorphic coding.

Security analysts widely use different techniques, tools and mechanisms to perform behavior analysis. One of the most widely used tools for malware behavior modeling is Cuckoo, although researchers also prefer tools like VirMon, and WINAPIOverride32 to do malware analysis. All these tools have their own strength and weakness that ultimately affect the feature engineering process and sometimes the distinct features of great importance are skipped because of inherent limitations of these tools, which result in the poor performance of classification systems. In this paper, we aim to compare the efficiency and efficacy of agent-based and agent-less sandboxes in terms of detecting sophisticated malware variants. We use the Cuckoo agent-based sandbox and the VMRay agent-less sandbox for behavioral modeling of malware and later on, we measure the effectiveness of these tools in terms of detecting malicious software in general and, more specifically, sophisticated malicious software.

The remainder of the paper is organized as follows. Section II Relevant work, Section III Proposed research work, IV Experimental results, In Section V Future work, VI Conclusion and in Section VII Acknowledgment.

## II. RELATED WORK

Malware detection, clustering, and classification represents also a hot topic for academic researchers and industry professionals, here we are presenting different approaches which have used so far for detecting malware. First, some methods, which use static features, are described and then dynamic features are explained. Authors of [7] used static analysis to detect malware and benign samples by extracting 4-gram features from portal executable and later on these features were used to distinguish a benign sample from the malicious sample. In a more recent study, Opcode was used to detect malicious files [8]. Malware samples were reverse engineered to get an opcode, which is part of machine language instruction and depict the operations to be performed. It plays important role in distinguishing the legitimate software from malicious software. In [9] the authors presented a model to detect malware variants on the base of the byte frequency. In this model, suspicious malware is detected if its byte frequency is similar to some known malware class. In [10], the focus is on API sequence, which appears to be more frequently in most of the malware files and then applied similarity measure for the sequence. In [11], researchers proposed a model to detect suspicious samples by calculating frequencies of features for e.g. Dynamic link libraries, APIs and PE header, then use the information to gain feature selection to mark samples as malware or benign. In [12], authors proposed a malware classification system based on an n-gram feature vector. Dynamic analysis was done to get network level artifacts and then from these artifacts, n-gram feature vector was constructed and later on used for classification of malware. Authors in this paper claim to achieve 80% accuracy in classification by using a number of machine learning algorithms. Researchers in [13], analysed the system states changes, such as the number of new processes created, file written etc. The proposed method was evaluated on unseen malware variants whose signature was not available at the time and used a tree structure based on single linkage clustering to measure similarity among the various groups of malware.

In [14], the authors tried to address a problem related to the packing of malware, a technique based on obfuscation method to hide malware code in software. To address this problem, authors propose a technique that generates the signature of every packed malware. The dataset for analysis was divided into two parts. The first part is used for constructing between different systems entities, which include processes, system register etc. and other part for testing and evaluation. Authors in [15] propose a detection system based on a quantitative data flow model and then use graphs to depict the communication between different systems entities, which include processes, system register etc. Researchers in [16] concluded that there are repetitive actions on data sequence that malware mostly do, such as loops performing decryption or encryption, and this can be addressed through iterative system calls pattern mining. In [17], it was assumed that behavior of each executable can be represented by the values of register contents in its run-time.

Researcher in [18] used 4-grams to model API call sequences. By comparison of the average confidence of all 4-grams, samples are classified as malware or benign class.

[19] introduced a model sample behavior based on 2-gram features through system calls and their arguments by using prioritizing arguments. It successfully identified novel classes of malware with similar behavior and assigning unknown malware to these discovered classes. Researchers in [20] removed the function libraries constructed by benign files from those which appeared in malware as segment threat, calculated segment entropy and extracted 3-grams Opcode for each segment.

In [21], in contrast to traditional techniques, the authors used iterative pattern mining to detect malware based on the assumption that malware do repetitive action on data sequence ranging from running infections to running loops, which perform a decryption/encryption process. The authors of the study break down the overall process into five steps. In the first step, they gathered malware samples and in next step, they captured PE interaction with operating system APIs by running these samples in the controlled virtual environment for e.g. VMWARE and Qemu are used. In this step API, call logs are used to construct the dataset and furthermore, in this, they see iterative API patterns that occur more than a minimum threshold. In the last step, they use pattern features as a dataset to classify malware. Different algorithms are used to train model for e.g. SVM, Random Forest etc. and they claim to achieve 95% accuracy with 98.4% detection rate.

## III. PROPOSED RESEARCH WORK

As summarized in section II, prior research led to capable platforms for dynamic analysis of malware. These are two types: Basic dynamic analysis and advance dynamic analysis platform. Dynamic analysis tools such as Capture-Bat, Regshot, APATE DNS, PEID, PE explorer, or Sysinternal were used to carry out the analysis, whereas advanced dynamic analysis tools like Virmon, Cuckoo, WINAPIOverride32 were used to model the behavior of malware. Most of the advanced dynamic analysis platforms are agent-based, and they usually drop their agent on the analysis machine to capture the features of malicious software, but current malware are intelligent enough to circumvent analysis when they find themselves being analyzed by the agent base sandbox and detonate themselves before being analyzed. In this research, we have taken both agent-based and agent-less sandbox and executed the malware sets in these environments in order to find which one is the best for capturing sophisticated malware features as shown in fig.1.

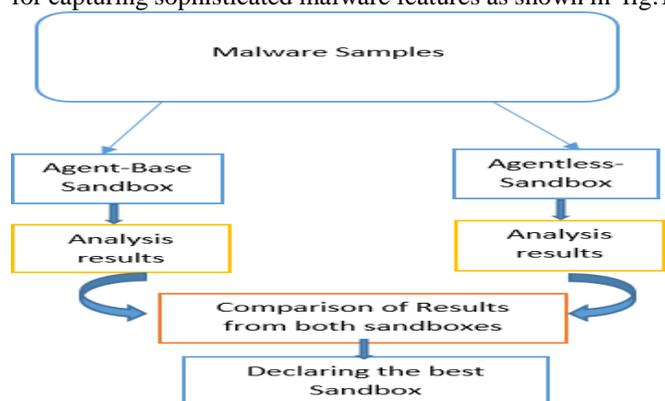

Fig 1. Conceptual model

TABLE I
COMPARISON OF PROPOSED MALWARE CLASSIFICATION METHODS WITH CURRENT STUDY.

| Author Name | Year | Technique used | Features | Representation |
|---|---|---|---|---|
| L. Bilge and T. Dumitras | 2014 | Dynamic analysis | communication between different system entities (processes, sockets, files or system registries) | Data Flow Graph |
| Mohaisen et al | 2014 | Dynamic analysis | N-gram feature of the network artifacts | N-gram |
| Ahmadi et al | 2013 | Dynamic analysis | executables' API call | Graph |
| Ghiasi et al | 2013 | Dynamic analysis | register values | The binary vector of Features |
| Ravi and Manoharan | 2012 | Dynamic analysis | API call sequences | N-gram |
| Rieck et al | 2011 | Dynamic analysis | system calls and their arguments | N-gram, Binary vector of Features |
| Tahan et al | 2012 | Dynamic analysis | API calls and their parameters | The binary vector of Features |
| Mansour Ahmadi et al | 2013 | Iterative pattern mining | API, call logs | API, call logs |

For conducting this research, we have used two different testbeds one with Cuckoo Sandbox and another with VMRay analyzer Sandbox. Cuckoo Sandbox is an open source widely used platform to model the behavior of malware in a controlled environment. It was developed as a summer project in 2010 in Google sponsored summer code project. The propose of this malware analysis system is to provide automatic analysis of malware for e.g. files created, deleted, API calls, argument and there return values etc. Cuckoo mainly focus on DLL, PDF, office documents, and different executables for windows and further consider Java files. VMRay Analyzer is agentless dynamic behavior analysis tool for malware. Unlike other established sandbox solutions in the market, it is embedded in the hypervisor in order to monitor the behavior of malware and overcome the problem in tradition sandboxes, thus malware could not able to detect that it is being detonated in control environment One reason for choosing VMRay analyzer is that it overcome the advance evasion techniques and another reason for choosing is its significant features for e.g. Evasion Resistance, Customizable Yet Automated, Easy Deployment, VMRay's Reputation Engine and Seamless Integration as shown in Table II.

In the first one, we used VMware workstation version (12.5.9) virtual environment, where we set up a virtual machine of Ubuntu (16.04 LTS) with Cuckoo sandbox installed on it to carry out the dynamic analysis of malware in order to get the artifacts of malicious software for understanding the behavior of malware. To execute malware in control environment in Ubuntu we have set up VBox with a Windows XP-SP3 machine, furthermore, the cuckoo agent was installed on the XP virtual machine along with some other software so that it can effectively capture the behavior of malicious variants when they are executed. In the second testbed, we have used VMRay analyzer, which was hosted on the VMRay cloud environment and we were been given special access by VMRay analyzer company for 30 days to perform our experiments. The testbed environment in the cloud was configured with almost all versions of Windows ranging from Windows XP (SP1, SP2, and SP3), Windows 7, Windows 8 and Windows 10 (with all service packs),

Moreover, in order to understand the behavior more clearly these machines were configured with all necessary software for e.g. MS Office 2007,2010,2013, Acrobat Reader version 9,10,12 etc. In our proposed conceptual model, we have used Cuckoo and VMRayanalyzer as shown in the fig. 2 and fig.3 to do a comparison of artifacts extracted by both the Sandboxes as shown in Table II.

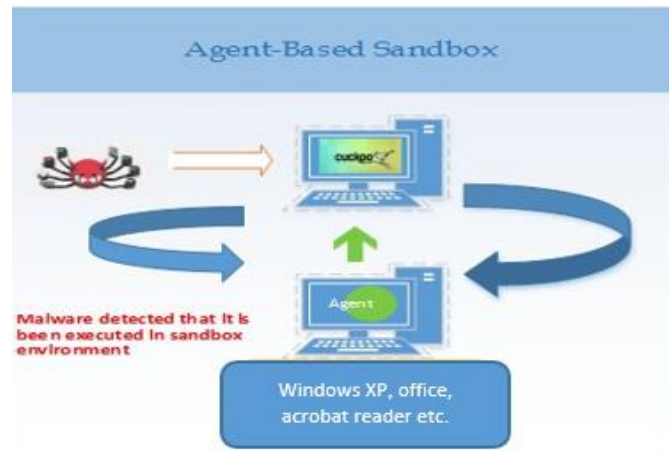
Fig. 2. Agent-based sandbox

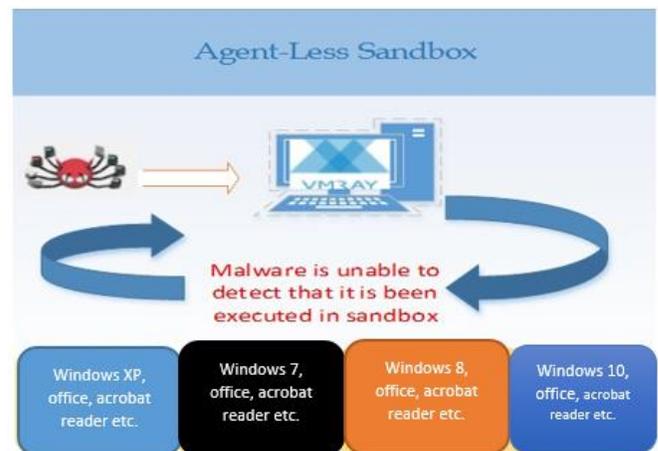
Fig. 3. Agent-less sandbox

TABLE II
DIFFERENCE BETWEEN CUCKOO AND VMRAY ANALYZER

| Sandbox | Zero Day Detection | Visualization of parent-child relationship | Evasion of Anti-analysis technique Possible or not? | Files related activities | Registry related activities | API calls |
|---|---|---|---|---|---|---|
| Cuckoo | × | × | × | √ | √ | √ |
| VMRay Analyzer | √ | √ | √ | √ | √ | √ |

## IV. EXPERIMENTAL RESULTS

We have taken malicious samples from different sources such as contagion dump and Zoo and executed these samples in both environments. Here we are mentioning the results of few of them, including The Ransomware Petya, Spyeye, Volatile Cedear, Dyre, and PAFISH. These samples were executed in both Agent-based and Agentless environment on a different version of Windows for e.g. Windows 7 (SP-1 32 bit and 64 bit), Windows 8(64bit) and Windows 10(64 bit) respectively as shown in the fig. 4. The behavior and actions of malicious samples were analyzed through dynamic analysis mechanisms using two different types of sandboxes: i) Agent-based sandbox and ii) Agentless sandbox tools. We performed empirical analysis by executing different samples of malware in both Agent-based and Agentless sandbox environments and find that agentless sandbox is more efficient in detecting sophisticated malware, which bypasses or crash themselves on finding themselves being detected by sandbox agent. The motivation behind this research is to model the behavior of those malware variants, which are able to thwart the agent base sandboxes and as a result, they are not detected.

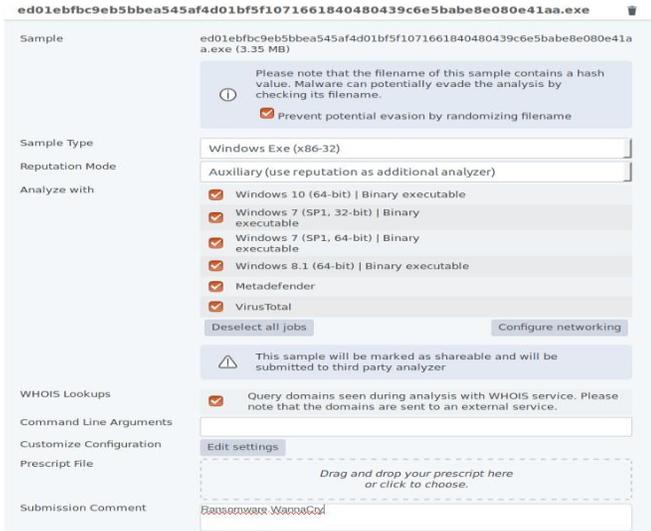

Fig. 4 Ransomware Wannacry

During the analysis, we closely monitored the changes occurring in the operating resource for e.g. DNS requests, HTTP requests, file related activities, registry related activities, API calls and their return values, service activities, IRC commands, and process tree we have seen during our experiment that whenever we try to execute some of the advance malware for e.g. Wanna Cry, Petya in Cuckoo it usually gives us very limited features and was unable to produce features which are helpful in understanding the behavior of malware.

In our view, the reason behind this is that almost of all advance malware usually monitor the running environment when they are executed and on finding the analysis platform they usually stop their execution or in some case give only insignificant features, whereas in contrast when these advance malware were executed in Agentless Sandbox the results were totally opposite and we were able to capture some of the features of significant importance for e.g. Zero-day detection, visualization of relationship between parent and child processes of malware etc. as shown in Table II, moreover we find that these features will play an important role in designing advance threat detection and mitigation platform and we will prove the significance of these features in our future work, so from our analysis we came to this conclusion that agentless sandboxes are more effective and robust in terms of capturing intelligent malware and in extracting their features.

## V. FUTURE WORK

This session discusses our future work and framework comprising three stages: the monitoring phase, feature-engineering phase, and learning stage.

### A. Monitoring stage

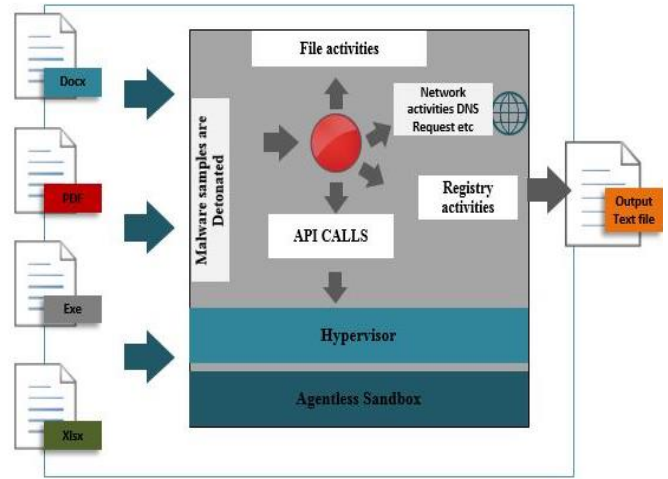

Fig. 5 Indigenous Sandbox for Dynamic analysis

In the monitoring stage, we will take malware samples from different classes and then will execute them to control the the environment in order to model the behavior of malicious software. For behavior analysis, we will use our indigenous in-house made agentless sandbox or some open-source Agentless sandbox as shown in the fig.5. and the reason behind using this tool is that we have found from experiments that Agentless Sandboxes are evasion resistance and nowadays sophisticated malware and APTs can detect it is being observed, therefore, as a result, they stop their executions. Also features used in [23] will be taken into consideration so we can collect as much information as possible from the execution of the malware.

### B. Feature Engineering stage

In this phase, feature sets will be created based on API calls and their argument as this is done by extracting string information from the text files generated by our indigenous tool or open source tool. Once we get the features, we will apply feature selection techniques to get significant feature sets and. In the last stage of this phase different NLP techniques for e.g. n-grams will be used to convert features into binary vectors, which are, later on, feed to deep learning algorithm for training purpose as shown in the fig. 6.

### C. Learning and verification stage

In this stage, the binary vector will be given to Learning algorithm as shown in fig. 6. In our case, we will use generative algorithms and our focus will be on deep belief network DBN and the reason behind using this is that they are very good at creating invariant representations of objects even if the specific object changes its size, contrast, angle etc. and they had produced very promising results in a number of image classification projects, moreover possess high accuracy. Moreover, the use of game theoretic models and trust will be explored [24].

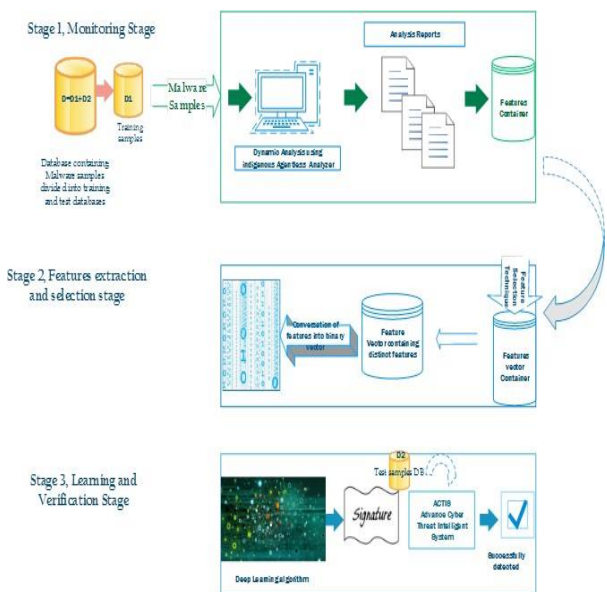

Fig. 6. Cyber Intelligent System for Malware detection

### VI. CONCLUSION

In this paper we reviewed past approaches for detecting malware through either static or through dynamic analysis, furthermore, we found that dynamic analysis is an effective approach for the behavioral analysis of malware. Dynamic analysis is usually carried out in sandbox environment, in our research we find that traditional sandboxes are not evasive resistance because they hook data by dropping their agent in control environment which can be detected by intelligent malware and as a result they don't unpack or execute themselves on finding agent, so in our research we find that Agentless Sandbox is the best solution for dynamic analysis.

### VII. ACKNOWLEDGMENT

This work was supported by CYBER-TRUST project, which has received funding from the European Union's Horizon 2020 research and innovation programme under grant agreement no. 786698.

REFERENCES

[1] K. H. Mathur and S., *A Survey on Techniques in Detection and Analyzing Malware Executables*, 2013, vol. 3.
[2] *"Internet security threat report*, 0. [Online]. Available: https://www.symantec.com/content/dam/symantec/docs/reports/istr-22-2017-en.pdf
[3] [Online]. Available: https://www.pandasecurity.com/mediacenter/pressreleases/malware-still-generated-rate-160000-new-samples-day-q22014/
[4] L. B. T. Dumitras, *Before we knew it: an empirical study of zero-day attacks in the real world* , 2012.
[5] "Modeling and Analysis on the Propagation Dynamics of Modern Email Malware," *IEEE Transaction on dependable and Secure Computing*, vol. 2014, pp. 11–4.
[6] L. B. T. Dumitras, *Before we knew it,* Raleigh, NC, USA: ed, 2012.
[7] A. Walenstein, D. J. Hefner, and J. Wichers, "Header information in malware families and impact on automated classifiers,," *ed*, pp. 15–22, 2010.
[8] D. Bilar, "Opcodes as predictor for malware," *International Journal of Electronic Security and Digital Forensics*, pp. 156–168, 2007.
[9] S. Yu, S. Zhou, L. Liu, R. Yang, and J. Luo, Malware variants identification based on byte frequency, 2010, vol. 2.
[10] "Malware detection using assembly and API call sequences," J. Comput. Virol, vol. 7, no. 2, pp. 107–119, 2011.
[11] U. Baldangombo, N. Jambaljav, and S. J. Horng, "A static malware detection system using data mining methods. arXiv Prepr. arXiv 1308.2831," 2013.
[12] "Mohaisen A , West AG , Mankin A , Alrawi O . Chatter: classifying malware families using system event ordering," 2014, pp. 283–91.
[13] B. M, O. J, A. J, M. MZ, J. F, and N. J, Automated classification and analysis of internet malware.
[14] S. Naval, SPADE: Signature Based Packer Detection, 2012.
[15] V. Walenstein, D. J. Hefner, and J. Wichers, "Header information in malware families and impact on automated classifiers," Nancy, France, 2010, pp. 15–22.
[16] "Malware detection by behavioural sequential patterns," Comput. Fraud and Secur, vol. 8, pp. 11–19, 2013.
[17] L. J and W. MP, "Malware detection with different voting schemes, COMPUSOFT," Int J Adv Comput Technol, vol. 2014, no. 3.
[18] "Malware detection using windows API sequence and machine learning," Int. J. Comput. Appl, vol. 43, no. 17, pp. 12–16, 2012.
[19] K. Rieck, P. Trinius, C. Willems, and T. Holz, "Automatic analysis of malware behavior using machine learning,," Journal of Computer Security, vol. 19, no. 4, pp. 639–668, 2011.
[20] R. Tian, R. Islam, L. Batten, and S. Versteeg, "Differentiating malware from clean ware using behavioral analysis," Nancy, 2010.
[21] M. Ahmadi, "Malware detection by behavioural sequential patterns."
[22] Computer Fraud & Security, vol. 2013, no. 8, pp. 11–19, 20
[23] K.-P. Grammatikakis, A. Ioannou, S. Shiaeles, and N. Kolokotronis, "Are cracked applications really free? an empirical analysis on Android devices," 16th IEEE International Conference on Dependable, Autonomic and Secure Computing — DASC, August 2018
[24] K. Ntemos, N. Kolokotronis, and N. Kalouptsidis, "Using trust to mitigate malicious and selfish behavior of autonomous agents in CRNs," in 27th IEEE Annual International Symposium on Personal, Indoor, and Mobile Radio Communications — PIMRC, September 2016, pp. 1–7.